**A general approach for analyzing baseline power spectral densities: Zwanzig-Mori projection operators and the generalized Langevin equation**

**Authors:** David Hsu[1], Murielle Hsu[1], John M. Beggs[2]

1. Department of Neurology, University of Wisconsin, Madison WI, USA
2. Department of Physics, Indiana University, Bloomington IN, USA

**Abstract**: There continues to be widespread interest in $1/f^{\alpha}$ behavior in baseline power spectral densities (PSD's) but its origins remain controversial. Zwanzig-Mori projection operators provide a rigorous, common starting place for building a theory of PSD's from the bottom up. In this approach, one separates out explicit "system" degrees of freedom (which are experimentally monitored) from all other implicit or "bath" degrees of freedom, and then one "projects" or integrates out all the implicit degrees of freedom. The result is the generalized Langevin equation. Within this formalism, the system PSD has a simple relation to the bath PSD. We explore how several models of the bath PSD affect the system PSD. We suggest that analyzing the baseline can yield valuable information on the bath. The Debye model of acoustic bath oscillations in particular gives rise to a low frequency $1/f$ divergence. Other power law behaviors are possible with other models. None of these models require self-organized criticality.

**Introduction**

Many complex systems have been observed to produce baseline power spectral densities (PSD) with a $1/f^{\alpha}$ power law form where $f$ is the frequency and where typically the exponent $\alpha$ is in the range $1 \leq \alpha \leq 2$. Such baseline spectra are observed in biological [1-4], physical [5, 6], psychological [7, 8], and economic systems [9], leading some authors to speculate that these spectra are signatures of a universal mechanism underlying dynamics in complex systems [10, 11]. Power law behavior of a given system property suggests that that property is scale-free, in the sense that changing the time scale or length scale associated with that property results in a function that also has a power law form, and that has exactly the same power law exponent. Scale-free systems exhibit patterns of activity that span all length and time scales, from the smallest and fastest to the largest and slowest. Such scale-free behavior is seen in systems at a thermodynamic critical point [12] and also in systems at self-organized criticality [13]. Self-organized criticality refers to systems that evolve towards a scale-free state on its own, without the need to tune thermodynamic variables such as the pressure and temperature in order to arrive at the scale-free state. These two forms of criticality are quite different [10, 11].

The ubiquity of $1/f^{\alpha}$ baseline PSD's has attracted a great deal of attention [14-16] because of the possibility that there may be a unifying underlying mathematical explanation. Although specific mathematical models have been shown to yield $1/f^{\alpha}$ baseline PSD's [17-20], a universal mathematical explanation of $1/f^{\alpha}$ baseline PSD behavior has proved elusive. Recent work has highlighted the fact that power law distributions do not automatically imply criticality [21]. In such non-critical systems, processes like combinations of exponentials [22], successive fractionation [23] and random walks [24] can generate power law distributions. These processes do not bring a system to the critical point, and do not imply scale-free interactions. There is now some skepticism that $1/f^{\alpha}$ baseline PSD behavior implies either thermodynamic or self-organized criticality [15].

In a system that exhibits $1/f^\alpha$ baseline PSD, is it possible to say whether that system is critical? If yes, can we say whether the type of criticality is the thermodynamic type or self-organized criticality? Are there examples of systems with $1/f^\alpha$ power spectra that really are critical? Should we pay *any* attention to $1/f^\alpha$ power spectra?

It would be desirable to take a first principles approach to understanding power law spectra by starting from fundamental physical laws and trying to derive an equation of motion that applies to all experimental systems, i.e., to seek a *universal equation of motion*. For systems that obey the laws of classical or quantum physics, we can safely take the laws of classical or quantum physics themselves to be the unifying determinant of system dynamics. However, in general these systems consist of an enormous number of microscopic degrees of freedom, of the order of the number of atoms or molecules in the system. It is not possible to monitor all of these degrees of freedom in real life experiments. Generally one monitors only a tiny fraction of the whole.

The problem of describing the dynamics of a few experimental variables when these variables are coupled to many other variables about which we know very little was addressed in a rigorous way independently by Zwanzig [25] and Mori [26]. The approach utilized projection operators, which allowed one to separate the experimental variables from all the other variables, and then to "integrate out" all the other variables, leaving one with an equation of motion for just the experimental variables. The result is the generalized Langevin equation (GLE). The generalized Langevin equation, as justified via Zwanzig-Mori projection operators, is thus the universal equation of motion that we seek.

Zwanzig-Mori projection operators and the GLE have over time proved extremely useful in nonequilibrium statistical mechanics [27-29]. However, there have been only a few applications outside of the statistical mechanics literature [30, 31]. In what follows, we investigate under what conditions the GLE, as a universal equation of motion, can give rise to power law baseline spectra.

**Theory**

Consider an experiment where a certain number of macroscopic physical observables $x(n,t)$ are monitored, $n = 1$ to $N$. We will refer to these variables as the "system" or *explicit* degrees of freedom. These may assembled into a single vector, $X(t)$, the components of which are the individual $x(n,t)$. All the degrees of freedom that are not explicitly monitored in the experiment, but that may interact with or influence the explicit degrees of freedom, are referred to as the "bath" or *implicit* degrees of freedom. The number of implicit degrees of freedom, for a general macroscopic system, may be exceedingly large. Nonetheless, if at a microscopic level, all the explicit and implicit degrees of freedom obey the laws of classical (or quantum) physics, then the equation of motion for $X(t)$ can be written in the form of a generalized Langevin equation:

$$\ddot{X}(t) = -K \circ X(t) - G \circ \dot{X}(t) + F_R(t)$$
$$\equiv -\int_0^{t-t_0} dt' \, K(t-t') X(t) - \int_0^{t-t_0} dt' \, G(t-t') \dot{X}(t') + F_R(t) \qquad \text{Eq (1)}$$

Here $t_0$ is the initial time in some arbitrary time window. The derivation of Eq (1) is most easily accomplished using Zwanzig-Mori projection operators [27]. We do not repeat the derivation here; also see the discussion in Ref [32]. The open circles in Eq (1) denote time convolutions. We have normalized $X(t)$ such that the time average of every vector component of $X(t)$ is zero. Formal expressions for $K(t)$, $G(t)$ and $F_R(t)$ in terms of the microscopic degrees of freedom are well known [27, 28]. In what follows, we shall not need them. We only need to know the form of the equation of motion, Eq (1).

Equation (1) represents a universal equation of motion, in that it applies to all macroscopic degrees of freedom and any conceivable experiment as long as (1) the total system obeys the classical or quantum laws of physics, (2) the variables $X(t)$ and $\dot{X}(t)$ are independent degrees of freedom, and (3) the system is stationary [32]. In particular, Eq (1) is not restricted to Brownian motion, nor to linear dynamics, nor to equilibrium or ergodic dynamics [27]. The matrix $K(t)$ in Eq (1) plays the role of a time-delayed force constant, which describes how strongly the explicit degrees of freedom interact with one another, and at what time scale. The matrix $G(t)$ plays the role of a time-delayed friction, which causes energy in the explicit degrees of freedom to dissipate into the "bath" degrees of freedom. It is often referred to as the friction or memory kernel. The vector $F_R(t)$ plays the role of a "random force", which represents the force directly exerted on the explicit system degrees of freedom by the implicit bath degrees of freedom.

Let us define pair correlation functions as follows:

$$C_{AB}(t) = \int_{-\infty}^{\infty} dt_0 \, A(t+t_0) B(t_0) \equiv <A(t)B(0)> \qquad \text{Eq (2)}$$

where A(t) and B(t) can represent $X(t)$, $\dot{X}(t)$ or $F_R(t)$. If the variables $X(t)$ and $\dot{X}(t)$ are independent degrees of freedom, then one should find that the time average of the product of $X(t)$ and $\dot{X}(t)$ should equal the product of the individual time averages, which in turn by construction are each equal to zero: $C_{XV}(0) = <X(0)><\dot{X}(0)> = 0$. On the other hand, if $X(t)$ and $\dot{X}(t)$ are not independent degrees of freedom, then one will find that $C_{XV}(0) \neq 0$ even if averaged over long times. In this case, rather than using Eq (1), one may use a more general form of the generalized Langevin equation:

$$\dot{X}(t) = -W \circ X(t) + F_R(t) \qquad \text{Eq (3)}$$

In what follows, we will assume that $X(t)$ and $\dot{X}(t)$ are indeed independent degrees of freedom.

By stationary dynamics is meant that time averaged properties of the system remain constant if averaged over a sufficiently long period of time. A hallmark of a stationary system is the Wiener-Khinchin theorem, which states that the Fourier transform of the autocorrelation function of any degree of freedom yields its power spectral density (PSD). If the Fourier transform of $C_{AB}(t)$ is written as

$$\hat{C}_{AB}(\omega) = \int_{-\infty}^{\infty} dt \exp(-i\omega t) \, C_{AB}(t) \qquad \text{Eq (4)}$$

then stationarity implies that

$$\hat{C}_{AB}(\omega) = \hat{A}(\omega)\hat{B}(\omega)^* \equiv S_{AB}(\omega).  \quad\quad\quad\text{Eq (5)}$$

Here the asterisk implies complex conjugation. The diagonal elements of $S_{AB}(\omega)$ are the respective power spectral densities, e.g., $S_{AA}(\omega)$ is the power spectral density of $A(t)$.

Note that stationarity does not imply that the system is linear, nor is it free of fluctuations. However, an important consequence of stationarity is that fluctuations induced by the random force must be matched over time by dissipative forces acting through the friction term. This concept is known as the fluctuation-dissipation theorem [33]. For Eq (1), the fluctuation-dissipation theorem implies that the time correlation of the random force with itself is proportional to the friction kernel. This relationship may be stated equivalently either in the time or frequency domains:

$$C_{RR}(t) = \langle F_R(t) F_R(t_0) \rangle = C_{VV}(0) G(t) \quad\quad\quad\text{Eq (6a)}$$

$$\hat{C}_{RR}(\omega) = C_{VV}(0) \hat{G}(\omega) = S_{RR}(\omega) \quad\quad\quad\text{Eq (6b)}$$

Note that Eq (6b) states that the Fourier transform of the friction kernel is proportional to the power spectral density of the random force. We shall need these equations later.

Another important concept is that the random force, by definition, should not be correlated with any of the explicit degrees of freedom [27], else it would not truly be a random force. This lack of correlation between the random force and the explicit degrees of freedom can be utilized as the basis of a *variational principle* by which $K(t)$, $G(t)$ and $F_R(t)$ may be extracted from experimental data [32].

In this paper, we wish to calculate the baseline power spectral density (PSD) of a single explicit degree of freedom, neglecting the interaction of this degree of freedom with any other explicit degree of freedom. Therefore, we shall ignore the term in $K(t)$ in

Eq (1). Let $X(t)$ now refer to a single degree of freedom. If the Fourier transform of $X(t)$ is written as

$$\hat{X}(\omega) = \int_{-\infty}^{\infty} dt \exp(-i\omega t) X(t) \qquad \text{Eq (7)}$$

then the PSD of $X(t)$ is given by:

$$S_{XX}(\omega) = \hat{X}(\omega)\hat{X}(\omega)^* \qquad \text{Eq (8)}$$

If one defines the time correlation function of $X(t)$ as

$$C_{XX}(t) = \int_{-\infty}^{\infty} dt_0\, X(t+t_0)X(t_0) \equiv <X(t)X(0)> \qquad \text{Eq (9)}$$

then stationarity (see Eq 5) implies:

$$S_{XX}(\omega) = \hat{C}_{XX}(\omega) \qquad \text{Eq (10)}$$

Similarly, if one defines a velocity $V(t) = \dot{X}(t)$, then one may define a velocity correlation function as:

$$C_{VV}(t) = \int_{-\infty}^{\infty} dt_0\, V(t+t_0)V(t_0) \equiv <V(t)V(0)> \qquad \text{Eq (11)}$$

The Fourier transform of $V(t)$ is related to $\hat{X}(\omega)$ through $\hat{V}(\omega) = -i\omega\hat{X}(\omega)$. Putting together these various relations, one finds that

$$S_{XX}(\omega) = \frac{\hat{C}_{VV}(\omega)}{\omega^2} \qquad \text{Eq (12)}$$

An equation of motion for the velocity correlation function can be derived from Eq (1) by evaluating Eq (1) at time $t+t_0$, multiplying both sides by $V(t_0)$ and integrating over all $t_0$:

$$\dot{C}_{VV}(t) = -G \circ C_{VV}(t) \qquad \text{Eq (13)}$$

In Eq (13), because the random force should not be correlated with the explicit degree of freedom, we have used $\langle V(t)F_R(0)\rangle = 0$.

Let $\tilde{C}_{VV}(z)$ represent the one-sided Fourier-Laplace transform of $C_{VV}(t)$:

$$\tilde{C}_{VV}(z) = \int_0^\infty dt\, e^{-zt} C_{VV}(t) \qquad \text{Eq (14)}$$

Taking the Fourier-Laplace transform on both sides of Eq (13) and solving for $\tilde{C}_{VV}(z)$ then yields:

$$\tilde{C}_{VV}(z) = \frac{C_{VV}(0)}{z + \tilde{G}(z)} \qquad \text{Eq (15)}$$

where we will take $z = i\omega$. The Fourier transform of $C_{VV}(t)$, $\hat{C}_{VV}(\omega)$, is related to the Fourier-Laplace transform $\tilde{C}_{VV}(i\omega)$ by

$$\hat{C}_{VV}(\omega) = \tilde{C}_{VV}(i\omega) + \tilde{C}_{VV}(i\omega)^* = 2\operatorname{Re}\{\tilde{C}_{VV}(i\omega)\} \qquad \text{Eq (16)}$$

Thus the PSD of $X(t)$ is given by:

$$S_{XX}(\omega) = \frac{2C_{VV}(0)}{\omega^2} \text{Re}\left\{\frac{1}{i\omega + \tilde{G}(i\omega)}\right\}$$

$$= \frac{2C_{VV}(0)}{\omega^2}\left\{\frac{G_R(\omega)}{G_R(\omega)^2 + [\omega + G_I(\omega)]^2}\right\}$$

Eq (17)

where $G_R(\omega)$ and $G_I(\omega)$ are the real and imaginary parts of $\tilde{G}(i\omega)$.

We now have an expression for the baseline system PSD in terms of the Fourier-Laplace transform of the friction kernel $G(t)$. Equation (17) is our main theoretical result. Note that, in analogy with Eq (16) and referencing Eq (6b), $G_R(\omega)$ is in fact proportional to the PSD of the bath: $G_R(\omega) = S_{RR}(\omega)/2$. Thus, Eq (17) relates the baseline system PSD in a simple way to the bath PSD.

Thus far, in deriving Eq (17), we have made no approximations aside from dropping the direct interactions between the explicit degrees of freedom by dropping the term in $K$ in Eq (1). This step was necessary in separating out the baseline PSD from the total PSD. Now we need an approximate expression for the friction kernel $G(t)$. It is already clear from Eq (17) that a great variety of frequency dependences are possible for the baseline PSD, depending on the frequency spectrum of the friction kernel. We shall next consider 3 simple illustrative examples.

**Examples**

*Example: Brownian motion, $G(t) = \lambda\delta(t)$.* In this example, the random force autocorrelation function decays instantaneously, and the friction kernel has no "memory" of what came before. This kind of friction is called Markovian friction. The Fourier-Laplace transform of $G(t)$ is given by $\tilde{G}(i\omega) = \lambda/2 = G_R(\omega)$, which implies that $S_{RR}(\omega) = \lambda$. Therefore, as is well-known, this model consists of "white noise" bath, with equal random force contributions from all frequencies. The system PSD is then given by

$$S_{XX}(\omega) = \frac{C_{VV}(0)\lambda}{\omega^2\left[\omega^2 + (\lambda/2)^2\right]} \qquad \text{Eq (18)}$$

The system PSD therefore goes as

$$S_{XX}(\omega) \approx \frac{4C_{VV}(0)}{\lambda\omega^2} \qquad \text{for } \omega \ll \lambda, \qquad \text{Eq (19a)}$$

$$S_{XX}(\omega) \approx \frac{C_{VV}(0)\lambda}{\omega^4} \qquad \text{for } \omega \gg \lambda, \qquad \text{Eq (19b)}$$

yielding the well-known power laws with $\alpha = 2$ and $\alpha = 4$ in the two different frequency regimes. From experimental plots of $S_{XX}(\omega)$ spanning both regimes, one can also easily extract the coupling strength $\lambda$.

*Example: Debye acoustic phonon bath.* When a great many molecules condense into a crystalline solid, large scale collective motions spanning the entire solid may arise which at low enough temperatures can be described as nearly harmonic oscillators [34]. These oscillations are known as phonons. The force exerted by a phonon mode on any coupled degree of freedom is linearly proportional to the displacement of the phonon coordinate from equilibrium, which in turn oscillates sinusoidally. Thus one may write the friction kernel as:

$$G(t) = \int_0^\infty d\omega\, D(\omega)\lambda(\omega)\cos(\omega t) \qquad \text{Eq (20)}$$

Here $\omega$ is the frequency of a phonon mode, $\lambda(\omega)$ characterizes how strongly this phonon mode is coupled to the system degree of freedom, and $D(\omega)$ characterizes the number of phonon modes of frequency $\omega$, i.e., it is the phonon *density of states*. Taking the Fourier-Laplace transform of Eq (20) yields:

$$\tilde{G}(i\omega) = \frac{\pi}{2} D(\omega)\lambda(\omega) - i\int_0^\infty d\omega' D(\omega')\lambda(\omega') P\left[\frac{1}{\omega-\omega'} + \frac{1}{\omega+\omega'}\right] \quad \text{Eq (21)}$$

Here $P$ stands for a principal value. We then have

$$S_{XX}(\omega) = \frac{\pi C_{VV}(0)}{\omega^2} \left\{ \frac{D(\omega)\lambda(\omega)}{[\pi D(\omega)\lambda(\omega)/2]^2 + [\omega - \Omega(\omega)]^2} \right\} \quad \text{Eq (22)}$$

where

$$\Omega(\omega) = \int_0^\infty d\omega' D(\omega')\lambda(\omega') P\left[\frac{1}{\omega-\omega'} + \frac{1}{\omega+\omega'}\right] \quad \text{Eq (23)}$$

For an isotropic 3-dimensional solid, the phonon density of states increases as $\omega^2$ at frequencies lower than a characteristic frequency, known as the Debye frequency $\omega_D$. These phonons are referred to as acoustic phonons, and were introduced by Debye in his classic explanation of the low temperature behavior of the specific heat [35]. For continuously deformable media, the coupling strength can be approximated as being linear in $\omega$ [34]. This approximation is known as Ohmic dissipation. For $\omega \leq \omega_D$, we may therefore write $D(\omega)\lambda(\omega) = \lambda_0 [\omega/\omega_D]^3$.

If the coupling strength is weak, then one may ignore $\Omega(\omega)$ in Eq (22). Typically this term introduces no more than a frequency shift in the PSD. For convenience, define $\omega_0 = \sqrt{2\omega_D^3/(\pi\lambda_0)}$. The system PSD, for $\omega \leq \omega_D$, then becomes:

$$S_{XX}(\omega) = \frac{2 C_{VV}(0)}{\omega \omega_0^2 \left[1 + (\omega/\omega_0)^4\right]} \quad \text{Eq (24)}$$

The system PSD therefore goes as

$$S_{XX}(\omega) \approx \frac{2C_{VV}(0)}{\omega \omega_0^2} \qquad \text{for } \omega \ll \omega_0, \qquad \text{Eq (25a)}$$

$$S_{XX}(\omega) \approx \frac{2C_{VV}(0)\omega_0^2}{\omega^5} \qquad \text{for } \omega \gg \omega_0. \qquad \text{Eq (25b)}$$

Equations (25) predict a $1/f$ frequency dependence, without bound, as $f \to 0$, and a $1/f^5$ frequency dependence for larger $f$. A caveat is that the Debye acoustic phonon model breaks down at higher frequencies. No general formula accurately describes the density of states of the higher frequency vibrations of a solid, as these vibrations depend on the details on the intermolecular and intramolecular forces.

*Example: single damped harmonic oscillator, $G(t) = \lambda \exp(-\gamma t)\cos(\omega_0 t)$.* In this simple model, the bath consists of a single oscillator that damps with a decay rate of $\gamma$. Such a model may represent a localized vibration in a solid (i.e., an intramolecular or pseudolocal mode). We will take $\gamma < \omega_0$, so that the oscillator frequency is well defined. The Fourier-Laplace transform of $G(t)$ is then given by:

$$\tilde{G}(i\omega) = D(\omega) - i\Omega(\omega) \qquad \text{Eq (26)}$$

where

$$D(\omega) = \frac{\lambda\gamma/2}{(\omega-\omega_0)^2 + \gamma^2} + \frac{\lambda\gamma/2}{(\omega+\omega_0)^2 + \gamma^2} \qquad \text{Eq (27)}$$

$$\Omega(\omega) = \frac{\lambda(\omega-\omega_0)/2}{(\omega-\omega_0)^2 + \gamma^2} + \frac{\lambda(\omega+\omega_0)/2}{(\omega+\omega_0)^2 + \gamma^2} \qquad \text{Eq (28)}$$

The system PSD is then given by

$$S_{XX}(\omega) = \frac{2C_{VV}(0)}{\omega^2} \left\{ \frac{D(\omega)}{D(\omega)^2 + [\omega - \Omega(\omega)]^2} \right\} \qquad \text{Eq (29)}$$

Note that

$$D(\omega) \approx \frac{\lambda\gamma}{\omega_0^2 + \gamma^2} \qquad \text{for } \omega << \omega_0, \qquad \text{Eq (30)}$$

$$D(\omega) \approx \frac{\lambda\gamma}{\omega^2} \qquad \text{for } \omega >> \omega_0. \qquad \text{Eq (31)}$$

Note also that, at weak coupling, $\Omega(\omega)$ does not affect the system PSD except possibly by shifting the frequency scale by a little bit. We shall again ignore this term. The system PSD can then be written:

$$S_{XX}(\omega) = \frac{2C_{VV}(0)}{\omega^2 D(0)} \qquad \text{for } \omega << \omega_0, \omega << D(0) \qquad \text{Eq (32)}$$

$$S_{XX}(\omega) = \frac{2C_{VV}(0)D(0)}{\omega^4} \qquad \text{for } \omega << \omega_0, \omega >> D(0) \qquad \text{Eq (33)}$$

$$S_{XX}(\omega) = \frac{2C_{VV}(0)\lambda\gamma}{\omega^6} \qquad \text{for } \omega >> \omega_0, \omega >> D(0) \qquad \text{Eq (34)}$$

The system PSD thus shows power law behavior with 3 regimes, with power law exponents of 2, 4 and 6.

**Discussion**

The generalized Langevin equation, as justified through Zwanzig-Mori projection operators, is applicable to any macroscopic dynamics coupled to any kind of environmental bath. In principle it is applicable to the dynamics of anything from Brownian motion to earthquakes, from economics possibly even to human behavior. Using the generalized Langevin equation, we have shown that the baseline power spectral density of any observable property can be related to the power spectrum of the relevant bath states. Depending on what the bath consists of, a rich variety of system baseline power spectral densities are possible.

The Debye acoustic phonon bath model in particular yields a baseline behavior that goes as $1/f$ down to infinitesimal frequencies, with no bound. Might this model or something similar be applicable to bath states other than the collective vibration of crystalline solids? Nearly harmonic oscillations arise in many contexts as low energy states in coupled many-body systems. The reason is that low energy states tend to reside near the bottoms of potential energy wells. Because most potential energy wells in nature are continuous and well-behaved, the bottoms of these wells can often be approximated quite well as quadratic functions, thus yielding nearly harmonic oscillations when dynamics is confined to these regions.

When might the Debye model break down? The Debye model assumes an isotropic 3-dimensional solid, and it considers only the lowest frequency vibrational modes. For disordered systems with many trapped imperfections (e.g., glassy substances), the Debye model is likely to break down, although acoustic modes are still present [36, 37]. Exactly how it breaks down and how the baseline PSD is affected would be of interest, because it would teach us something about the bath states. In this respect, analysis of baseline PSD may potentially be useful even when $1/f^\alpha$ power law behavior is absent. The system variables may then be considered as probes that teach us something about the bath (see Eq 15). There is always something to be learned about bath states and bath dynamics.

In addition, Eq (15) suggests that it may be more natural to study the velocity correlation function $C_{VV}(t)$ of a given system variable rather than the more conventional $C_{XX}(t)$, because $C_{VV}(t)$ is more directly related to the bath correlation function, see Eq (6a). It is also a simple matter to show that $\hat{C}_{VV}(\omega)$ is well behaved in the three examples presented, whereas $S_{XX}(\omega) = \hat{C}_{XX}(\omega)$ diverges as either $1/f$ or $1/f^2$ in the low frequency limit. The divergence in $\hat{C}_{XX}(\omega)$ at low frequencies merely reflects the fact that the system variable $X(t)$ in these examples is not spatially bound (recall that we set $K(t)=0$ in order to isolate the baseline PSD from the total PSD). For example, the mean square displacement $\langle |x(t)-x(0)|^2 \rangle$ for the Brownian particle diverges linearly in time, $\langle |x(t)-x(0)|^2 \rangle = 2Dt$, where $D$ is the diffusion constant. This divergence at long times in the time domain corresponds to the $1/f^2$ divergence at small frequencies in the complementary Fourier transform frequency domain.

*Self-organized criticality and the thermodynamic critical point*

The classical model of self-organized criticality is the sandpile model. Here grains of sand are randomly dropped onto a sandy landscape in a sandbox [10, 11]. When a sandpile reaches a critical height above its neighbors, then it is allowed to topple by specific rules. The spatial and temporal distribution of toppling events are then found to obey power laws, meaning that topping events can involve a single grain of sand, or a few grains of sand up to many grains of sand covering the entire surface of the sandbox.

In this model, the "bath" is represented by infinitesimal spontaneous fluctuations in the location and velocity of individual grains of sand and very strong dissipation, such that any kinetic energy gained by a grain of sand during a toppling event is instantaneously sucked away, as soon as the grain of sand reaches its new location. Because new grains of sand are continuously added to the sandbox, the sandy landscape reaches a state such that it stays at the highest energy metastable state. From this metastable state, as more energy continues to be added incrementally, the system eventually escapes. It is because the system is maintained at the highest energy metastable state that the system is able to visit any other neighboring metastable state. Note that maintaining the system at this highest energy metastable state requires external input of energy. This model also assumes that dissipation is very strong, so that as soon as the system escapes from a metastable state, it is "quenched" into a zero kinetic energy state. This instantaneous quenching amounts to assuming that the bath time scales are much faster than the system time scales, and that the bath is very "cold."

The thermodynamic critical state is quite different. It does not require external input of energy but it does require thermodynamic variables such as the temperature and pressure to be "tuned" to the critical values. When tuned to a thermodynamic critical point, the system is able to hop from metastable state to metastable due to spontaneous fluctuations induced by random interactions with the bath. The magnitude of the bath fluctuations varies over a range that is wide enough that it can kick the system out of any metastable state. The system does not stay long in any single metastable state, and no metastable state is totally inaccessible to it. Furthermore, at the thermodynamic critical point, the time scale of bath relaxation does not have to be much faster than the system time scales. In fact, these time scales could overlap. Thus, quenching events are rare and the kinetic energy of the system is nearly always greater than zero. The system is nearly always moving in phase space, either rattling around inside a metastable energy well or crossing over an energy barrier and headed towards another energy well.

Neither type of criticality clearly describes the three examples we present, yet all three examples exhibit power laws in certain ranges. For self-organizing criticality, one would expect to see an external source of energy. One might argue that turning up the temperature in our examples (and hence the spontaneous bath fluctuations) may play the same role as an external source of energy, and that if one were to turn the temperature up high enough, the system energy can be increased right to the level of the highest metastable state, or even higher. But there is nothing in our examples that plays the role of this thermostat. The power laws depend only on the frequency regime, with no clear critical temperature below which the power laws disappear. One might also argue that the power law in self-organized criticality should extend over all frequency regimes, not piece-wise as in our examples with different power laws at different frequencies. However, in self-organized criticality, one is allowed to assume that the bath relaxes much faster than the system, and the frequency $f$ is restricted by this requirement. For the Debye phonon model, we can similarly restrict attention to frequencies $f \ll f_0$, and in this case we do find $1/f$ baseline PSD behavior for all frequencies within this regime.

Similarly, in comparison to the thermodynamic critical point, there is again no need in our examples to tune thermodynamic variables such as pressure and temperature to critical values. Thus it seems rather easy to produce examples where power laws are present but where there is no clear connection to a thermodynamic or self-organized critical point.

We comment that, aside from the Debye phonon bath, there is one other way to produce $1/f$ divergence of the system PSD at low frequencies, namely, when the system is coupled to a bath that also has $1/f$ divergence at low frequencies. This situation may be quite interesting. In this case, both system and bath degrees of freedom have long time tails that go like $1/t$. There is no separation of time scales between system and bath. This power law exponent is also the only one where the system and the bath have the same exponent (at low frequencies). Hence, the bath and system variables in this case are similar in some sense, and the distinction between *explicit* and *implicit* degrees of freedom is blurred. Even more interesting is that this system should, within the low frequency regime, be scale free, and that the scale free nature holds not just for the activation of system variables but it extends also to the bath variables and in exactly the same way, with the same power law dependence.

Therefore, the case where both the system and bath PSD's have a $1/f$ divergence at low frequencies may also represent a critical point. Let us refer to such systems as double $1/f$ systems. We are not certain if it corresponds to the thermodynamic critical point, but because there is no separation of time scales, it cannot correspond to self-organized criticality.

Another point of interest is the possibility of a hierarchy of bath vs system variables. For example, consider 3 sets of distinct degrees of freedom, A, B and C. Suppose that set A consists of Debye-like phonon modes that serve as a bath for set B. Then set B will have a $1/f$ divergence at low frequencies. Now suppose that set B serves as a bath for set C. Then set C will also have a $1/f$ divergence at low frequencies, and sets B and C form a double $1/f$ system. Therefore, it may be possible to create experimental double $1/f$ systems.

**Conclusion**

Using a general approach to studying system-bath sets, we find that it is rather easy to produce power law behavior in system power spectral densities without invoking either thermodynamic criticality or self-organized criticality. Thus power laws in PSD's in themselves do not necessarily imply self-organized or thermodynamic criticality. Nonetheless, one can regard baseline PSD power laws as a kind of signature of the coupled bath, because if different power laws are seen in different frequency regimes, one may be able to determine what kind of bath is coupled to the system variables. Thus studying the baseline PSD may provide valuable information about the bath.

In our analysis, we have assumed that the explicit degrees of freedom do not significantly interact with each other, which allowed us to ignore the term in $K(t)$ in Eq (1). If there is significant interaction between explicit degrees of freedom, on the other hand, then one will need to treat these interactions on an equal basis with interactions with the bath. A variational approach to this problem is described in Ref [32].

**Acknowledgements:** DH was supported by the NIH Loan Repayment Program. We thank Drs. Hiroshi Fujisaki, Tom Keyes, John Straub and Gregory Worrell for helpful discussions.